\documentclass[ba, preprint]{imsart}

\RequirePackage{amsthm,amsmath,amsfonts,amssymb}
\RequirePackage[numbers]{natbib}
\RequirePackage[colorlinks,citecolor=blue,urlcolor=blue,backref=page,backref=page]{hyperref}
\RequirePackage{graphicx}
\usepackage{array}

\usepackage{algorithm}
\usepackage{algpseudocode}

\usepackage{multirow}
\usepackage{caption}
\usepackage{subcaption}

\pubyear{2025}
\arxiv{00000}
\volume{TBA}
\issue{TBA}
\firstpage{1}
\lastpage{1}

\startlocaldefs
\theoremstyle{plain}

\theoremstyle{definition}

\theoremstyle{remark}

\DeclareMathOperator*{\argmin}{arg\,min}

\newcolumntype{L}{>{\centering\arraybackslash}p{.3\linewidth}}
\newcolumntype{K}{>{\centering\arraybackslash}p{.10\linewidth}}
\newcolumntype{J}{>{\centering\arraybackslash}p{.10\linewidth}}

\newcommand*{\mline}[1]{%
\begingroup
    \renewcommand*{\arraystretch}{1.1}%
   \begin{tabular}[c]{@{}>{\centering\arraybackslash}p{\linewidth}@{}}#1\end{tabular}%
  \endgroup
}

\usepackage{tcolorbox}
\tcbuselibrary{theorems}

\newtcbtheorem{idProp}{Ideal Property}
{colback=white,colframe=black,fonttitle=\bfseries}{th}

\endlocaldefs

\begin{document}

\begin{frontmatter}
\title{The Traceplot Thickens: Developing All-Purpose Convergence Diagnostics for any Markov Chain Monte Carlo Algorithm}
\runtitle{The Traceplot Thickens}
%\thankstext{T1}{A sample additional note to the title.}

\begin{aug}
%%%%%%%%%%%%%%%%%%%%%%%%%%%%%%%%%%%%%%%%%%%%%%%
%% Only one address is permitted per author. %%
%% Only division, organization and e-mail is %%
%% included in the address.                  %%
%% Additional information can be included in %%
\author[A]{\fnms{Luke}~\snm{Duttweiler}\ead[label=e1]{lduttweiler@hsph.harvard.edu}},
\author[B]{\fnms{Jonathan}~\snm{Klus,}\ead[label=e2]{jonathan\_klus@urmc.rochester.edu}}
\author[A]{\fnms{Brent A.}~\snm{Coull,}\ead[label=e3]{bcoull@hsph.harvard.edu}}
\author[C]{\fnms{Ruth J.}~\snm{Geller,}\ead[label=e5]{rgeller@bu.edu}}
\author[D]{\fnms{Birgit}~\snm{Claus Henn,}\ead[label=e6]{bclaus@bu.edu}}
\and
\author[B]{\fnms{Sally W.}~\snm{Thurston}\ead[label=e4]{sally\_thurston@urmc.rochester.edu}}
%%%%%%%%%%%%%%%%%%%%%%%%%%%%%%%%%%%%%%%%%%%%%%
%% Addresses                                %%
%%%%%%%%%%%%%%%%%%%%%%%%%%%%%%%%%%%%%%%%%%%%%%
\address[A]{Department of Biostatistics,
Harvard T.H. Chan School of Public Health\printead[presep={,\ }]{e1,e3}}

\address[B]{Department of Biostatistics and Computational Biology,
University of Rochester\printead[presep={,\ }]{e2,e4}}

\address[C]{Department of Epidemiology,
Boston University School of Public Health\printead[presep={,\ }]{e5}}

\address[D]{Department of Environmental Health,
Boston University School of Public Health\printead[presep={,\ }]{e6}}
\runauthor{L. Duttweiler et al.}
\end{aug}

\begin{abstract}
    Markov Chain Monte Carlo (MCMC) algorithms are frequently used to perform inference under a Bayesian modeling framework. Convergence diagnostics, such as traceplots, the Gelman-Rubin potential scale reduction factor, and effective sample size, are used to visualize and monitor how well the sampler has explored the parameter space and the mixing of multiple chains. However, these classic diagnostics can be undefined or ineffective when the sample space of the algorithm varies in dimension or has a large number of discrete parameters.  In this article, we develop a novel approach to produce convergence diagnostics in these difficult scenarios by mapping the original sample space to the real-line and then evaluating the convergence diagnostics on the mapped values. The effectiveness of our method is demonstrated on a MCMC algorithm sampling from a Dirichlet process mixture model. The proposed diagnostics are also used to evaluate the performance of a Bayesian kernel machine regression model for estimating the health effect of multi-pollutant mixtures in the Study of Environment, Lifestyle, and Fibroids. Based on diagnostics for the latter dataset, we then explain how we modify the MCMC sampler to  improve convergence.
\end{abstract}

\begin{keyword}[class=MSC]
\kwd[Primary ]{62F15}
\kwd[; secondary ]{65C05}
\end{keyword}

\begin{keyword}
\kwd{Markov Chain Monte Carlo}
\kwd{Convergence Diagnostic}
\kwd{Traceplot}
\kwd{Gelman-Rubin Potential Scale Reduction Factor}
\kwd{Effective Sample Size}
\kwd{Bayesian Kernel Machine Regression}
\end{keyword}

\end{frontmatter}

\section{Introduction}

Convergence diagnostics for Markov Chain Monte Carlo (MCMC) methods are tools designed to assist investigators with selecting stopping times for an MCMC algorithm and detecting possible issues in algorithm convergence. A large number of general-use diagnostics have been proposed for MCMC algorithms that sample from univariate or multivariate continuous spaces \cite{cowles1996markov, roy2020convergence, moins2023use}, multi-modal spaces with a large number of parameters \cite{dixit2017mcmc} or spaces with categorical parameters \cite{deonovic2017convergence}. Among the most widely-applied diagnostics are the effective sample size (ESS), the Gelman-Rubin potential scale reduction factor (known as the Gelman-Rubin diagnostic, the PSRF, or $\hat{R}$), and the traceplot \cite{gelman1992inference, roy2020convergence}. These and other diagnostic measures may be easily applied in appropriate scenarios through one of the many software implementations available \cite{vstrumbelj2024past}.

While no post hoc empirical measure is able to `prove' in any formal sense that an MCMC algorithm has successfully converged, convergence diagnostics should consistently be applied alongside MCMC methods as they provide the best means available of detecting potential convergence issues. Unfortunately, these tools are frequently ignored in the development of novel and complex MCMC applications due to the difficulty in applying the diagnostics discussed above. Standard diagnostics can have several drawbacks, particularly when: (1) the number of parameters changes during sampling \cite{kontoyiannis2022bayesian, masotti2024general, gonen2019tutorial}, in which case they are undefined, (2) the model contains multiple binary or discrete parameters \cite{ranciati2024mixtures, he2024joint}, in which case they are typically ineffective, or (3) the number of parameters is large \cite{zhang2024modelling}, in which case they can be difficult to implement.

When diagnostics are desired in these difficult situations, researchers typically must develop convergence diagnostics tailored specifically for that problem or domain. Examples include MCMC literature on Bayesian phylogenetic tree analysis \cite{lanfear2016estimating, magee2024trustworthy, whidden2015quantifying}, entity resolution \cite{aleshin2024convergence}, and geoscience imaging \cite{somogyvari2020convergence}.

In the present article we propose a projection-based approach to diagnosing convergence in MCMC methods that can be quickly and easily applied in virtually any situation. While we sacrifice some theoretical guarantees for the ease and flexibility gained, we empirically demonstrate that our approach is successful in identifying convergence issues. Additionally, appropriate application of our technique can lead to identification of non-convergence that standard diagnostics fail to recognize.

The remainder of this article is organized as follows. Section \ref{sec: limitations} identifies current limitations in standard convergence diagnostic techniques. Section \ref{sec: ourApproach} presents our novel \textit{proximity-map} based approach for quickly developing convergence diagnostics for virtually any MCMC algorithm, and discusses general use considerations and recommendations. Section \ref{sec: workingExamples} provides two examples of use-cases for the proposed diagnostics. In the first we evaluate the convergence of an MCMC algorithm for a Dirichlet process mixture model. In the second, we discuss improvements to the algorithm for fitting a probit Bayesian kernel machine regression model, demonstrated by estimating the effects of a mixture of whole blood metals on the incidence of uterine fibroids in the Study of Environment, Lifestyle, and Fibroids (SELF) cohort. Finally, Section \ref{sec: discussion} provides a discussion the limitations of our method, available software, and future directions.

\section{Limitations in Current Diagnostics}\label{sec: limitations}

It is important to note that no approach to diagnosing MCMC convergence, including the approach proposed in this paper, will correctly identify non-convergence in every scenario. The theory underlying MCMC techniques guarantees convergence to the target distribution as long as infinite samples are drawn. As there are no equivalent guarantees for finite samples, diagnostics are useful as tools only for identifying non-convergence, not proving convergence. This issue is particularly exacerbated when an unknown target distribution is `bumpy' or multi-modal and the MCMC chains fail to thoroughly explore the parameter space. 

Thus, instead of providing definitive proof of convergence, convergence diagnostics can be thought of as serving two important purposes. The first is as a protection against non-convergence when non-convergence is detectable. Although this is not always possible, having diagnostics at hand to inform an investigator of a discernible failure to converge is very important. The second is to assist an investigator in determining which component of the MCMC algorithm is slowing down convergence so that the algorithm may be improved. For example, an investigator viewing traceplots in a model with a small number of continuous parameters can visually identify if one or more of the parameters may be slowing down the convergence of the model, and then take steps to adjust tuning parameters or the algorithm itself to improve performance. For models with a large or varying number of parameters, this process can be more difficult. 

With these goals for convergence diagnostics in mind, we turn to multiple limitations on current state-of-the-art diagnostics which inhibit their application or efficacy in particular scenarios. 

\subsection{Varying Dimensionality}

While the prototypical MCMC algorithm samples from a joint distribution of a fixed number of parameters, an increasing number of algorithms allow the number of parameters present in the model to change with each iteration. This flexibility can be introduced for many purposes, including allowing for model selection \cite{kontoyiannis2022bayesian}, identifying an unknown number of regions of interest \cite{masotti2024general}, adjusting model resolution \cite{somogyvari2020convergence}, or performing unsupervised clustering \cite{lalonde2020discovering, gonen2019tutorial}.

Typical uni- or multi-variate convergence diagnostics defined on a stable number of parameters sampled over multiple iterations are not suitable for parameters that are defined in some iterations and non-existent in others. In particular, this can make it very difficult to assess the convergence of the parameters that exist in some iterations and not in others. 

In some scenarios, convergence may still be examined (perhaps questionably) by running diagnostics only on those MCMC iterations that contained a particular set of parameters. For example, in a covariate selection context with three candidate covariates associated with parameters $\beta_1, \beta_2$ and $\beta_3$, the convergence of parameter $\beta_1$ could be examined by removing all MCMC iterations in which any one of $\beta_1, \beta_2,$ or $\beta_3$ is excluded from the model. The obvious flaw here is that the convergence diagnostics will no longer be accounting for the model selection portion of the algorithm, instead treating the MCMC draws as if they are from a fundamentally different MCMC algorithm that does not include model selection. 

Unfortunately, even this suboptimal solution is not available when parameter definitions are not consistent between iterations. Consider an MCMC algorithm designed to locate an unknown number of regions of interest (see \cite{masotti2024general} for example) in which, in each iteration if $K$ regions are located, then parameters $\mu_1, \dots, \mu_K$ represent the $K$ region centroids. As the number of located regions grows and shrinks, the definition of each particular $\mu_k$ no longer remains stable and overall model convergence becomes very difficult to assess without first assuming a particular value for $K$ and then performing additional processing \cite{stephens2000dealing}.

While diagnostics for particular MCMC algorithms with varying-dimensionality do exist, they require a significant investment in research hours to develop, and are frequently not portable to other models and algorithms. 

\subsection{Binary or Discrete Parameters}

Another class of MCMC algorithms that poses a unique challenge for convergence diagnostics is the set of MCMC algorithms with a large number of binary or discrete parameters. Such large collections of discrete parameters are often used in mixture models \cite{he2024joint, ranciati2024mixtures, gonen2019tutorial}, in network or tree models \cite{whidden2015quantifying, kuipers2017partition, suter2021bayesian}, or simply as indicators of covariate inclusion or exclusion \cite{bobb2015bayesian}.

Most diagnostics, including traceplots, ESS, and the PSRF, are designed to operate on continuous values. Thus, while these measures are defined for binary or discrete parameters, applying them in these cases can lead to strange or misleading results.

For example, consider a situation in which we are sampling from a target distribution that includes the binary parameter $\nu$, with $\text{Pr}(\nu = 1) = .999$. In this scenario, our MCMC algorithm will very rarely draw $\nu = 0$, and will remain at $\nu = 1$ for the vast majority of iterations. This repeated sampling of the exact same value will appear as extremely high autocorrelation and therefore give a rather low ESS, and will be visualized on a traceplot as an almost perfectly flat line. While either of these would be a problem for a continuous parameter, in the case of $\nu$, this is actually what we would expect to see if the sampler was working correctly. While an experienced investigator knows to expect this behavior and will not be surprised, the results can still obscure issues. When a traceplot contains large sections of a flat line in a continuous parameter, it is clear there is a problem with the sampler. But with a discrete parameter this type of traceplot may be indicative of a problem \textit{or} may simply be the natural result of sampling from a discrete space. 

Ultimately, this only becomes an issue when there are a large number of these binary or discrete parameters. With a handful, an investigator can look at each individually and make an informed decision based on multiple diagnostics and other portions of the model. However, when there are hundreds, or even thousands, of binary parameters to consider, identifying any that are causing slow convergence using the available diagnostics becomes an onerous or potentially impossible task. 

\subsection{A Note on Likelihood-Based Diagnostics}\label{subSec: liklihood}

A common solution to the difficulties with convergence diagnostics discussed above is to evaluate diagnostics on the model score, log-score, posterior value at, or likelihood of, each MCMC iteration (see for example \cite{friedman2003being, grzegorczyk2008improving}). In order to use this technique, an investigator simply evaluates the likelihood (or one of the related measures) at each MCMC iteration and uses these values in a traceplot, to calculate the PSRF, or to evaluate one of the other diagnostic statistics. Since the likelihood in a large model essentially exists on a continuous scale and there will always be one likelihood value regardless of the number of parameters included in the MCMC iteration, this approach bypasses the issues discussed above. 

While the likelihood-based diagnostic approach can certainly be used to identify non-convergence in some cases, and can in fact be seen as a specific use-case of the more general framework we propose below, it has several weaknesses. 

For a model with a large number of parameters, it is possible that the likelihood space is multi-modal. If the likelihood space has multiple modes of similar likelihood value, then diagnostics based on the likelihood will treat chains or iterations in these different modes as if they were very close to one another in the \textit{parameter space}, although they are simply close to one another in the \textit{likelihood space}. This, of course, can lead to likelihood-based diagnostics that fail to identify non-convergence, even when other diagnostic approaches easily identify an issue. While this is not a difficulty if the modes in the likelihood all yield different values, the shape of the likelihood space is almost certainly not known to the investigator ahead of time. 

Despite this flaw, in many situations likelihood-based diagnostics will provide important non-convergence information to an investigator, particularly when used with multiple chains. However, projecting all MCMC iterations into the likelihood space obscures information about the individual parameters to a degree that, while non-convergence may be identified, the diagnostics will provide essentially no information to the investigator about the underlying \textit{cause} of non-convergence. 

As discussed at the beginning of Section \ref{sec: limitations}, diagnostics are not only used to identify non-convergence, but also to assist an investigator in identifying portions of the algorithm or particular parameters that may be a limiting factor in allowing the MCMC to converge. Projecting each MCMC iteration into the likelihood space eliminates information about the individual parameters, making these diagnostics significantly less useful for identifying the potential problems in an algorithm.

Likelihood-based diagnostics can be seen as a specific use-case of the more general framework we propose below. We certainly do not mean to discourage their use in general as they have been, and continue to be, informative tools when working with MCMC algorithms. Rather, we suggest including them as a part of a broader set of tools.

\section{The Proximity-Map Approach}\label{sec: ourApproach}

We now present a general approach for developing convergence diagnostics for almost any MCMC algorithm, including those with varying dimensionality or a large number of binary or discrete parameters. This approach is flexible, interpretable, and may be implemented quickly. In this section, the approach is described in general terms, with full working examples presented in Section \ref{sec: workingExamples}.

\subsection{Notation}

Let $\Theta$ be a parameter space on which an MCMC algorithm $\mathcal{M}$ is defined. Define $\mathcal{C} = \{C_1, \dots, C_K\}$ to be the full set of $k$ chains sampled from $\Theta$ using $\mathcal{M}$ and let $C_k = (\theta_{1k}, \dots, \theta_{nk})$ be a single MCMC chain with $n$ iterations sampled from $\Theta$ using $\mathcal{M}$, so that $\theta_{ik} \in \Theta$ for all $i,k$. In order to accommodate algorithms with varying-dimensionality we do not require each $\theta_{ik}$ to have a set dimension. 

We now define a \textit{proximity-map} to be a function $\mathcal{P}:\Theta \rightarrow \mathbb{R}$, which may depend on $\mathcal{C}$. While the proximity-map $\mathcal{P}$ does function in some ways like a standard projection, mapping a possibly large-dimensional space to the real-line, we choose to use the term proximity-map for two reasons. First, when using $\mathcal{P}$ we seek to preserve a meaningful sense of `distance' or `proximity' (discussed in detail below) during the mapping, hence proximity-map. Second, the term `projection' has a long, well-defined meaning in mathematics and we want to avoid any confusion in our definition.

In order to save space in some portions of the manuscript, we will slightly abuse notation and set $\mathcal{P}(C_k) = \big(\mathcal{P}(\theta_{1k}), \dots, \mathcal{P}(\theta_{nk})\big)$, and $\mathcal{P}(\mathcal{C}) = \{\mathcal{P}(C_1), \dots, \mathcal{P}(C_K)\}.$

Finally, we define $d:\Theta \times \Theta \rightarrow \mathbb{R}$ to be a \textit{pairwise distance} between draws from $\Theta$. For any $\theta_i, \theta_j \in \Theta$ we require:

\begin{itemize}
    \item $d(\theta_i, \theta_i) = 0$, and 
    \item $d(\theta_i, \theta_j) = d(\theta_j, \theta_i)$.
\end{itemize}

\noindent Note that our definition of distance differs from the common mathematical definition of a \textit{metric} in that we \textbf{do not} require $d(\theta_i, \theta_j) = 0 \Rightarrow \theta_i = \theta_j$, nor do we require the typical triangle inequality.

\subsection{Diagnostics based on a Proximity-Map}

The approach we now present is very simple in theory, and generally easy to implement in practice. However, like many aspects of utilizing MCMC algorithms, effective execution is more art than science. Thus, after presenting the approach we will spend the remainder of the paper discussing principles for good use of the framework, providing options for proximity-maps and pairwise distance functions, and demonstrating effective use through several detailed illustrations.

The proximity-map approach to convergence diagnostics can be performed as follows:

\begin{enumerate}
    \item Draw the $K$ chains $\mathcal{C}$ from the parameter space $\Theta$ using the MCMC algorithm $\mathcal{M}$.
    \item Select a proximity-map $\mathcal{P}:\Theta \rightarrow \mathbb{R}$.
    \item Evaluate $\mathcal{P}(\mathcal{C})$, returning a set of $K$ univariate chains.
    \item Run convergence diagnostics on the $K$ univariate chains. 
\end{enumerate}
\noindent While this approach appears simple, it is clear that step (2), selecting a useful and informative proximity-map $\mathcal{P}$, is crucial for the success of the method. In light of the importance of selecting a proximity-map, we will focus the remainder of this manuscript on guidelines and examples for the selection process, along with illustrating the effectiveness of approach when used correctly. 

\subsection{Ideal Properties for Proximity-Maps}

In general, there are two key `ideal properties' to consider when choosing a proximity-map $\mathcal{P}$. We refer to these as `ideal properties' because the requirements they impose are frequently not fully achievable no matter how the proximity-map is chosen. For explanations in this section, let $\theta_i, \theta_j, \theta_k \in \Theta$ be three possible realizations of the parameter vectors, with possible varying dimensionality. 

First, the proximity-map should accurately reflect a meaningful sense of `proximity', or distance, in the MCMC algorithm. Commonly used convergence diagnostics such as the traceplot, the effective sample size, and the Gelman-Rubin PSRF operate with a Euclidean understanding of proximity. That is, since they are based on Euclidean measures like auto-correlation and variance, these diagnostics are only effective when we can safely \textit{assume} that a smaller Euclidean distance between $\theta_i$ and $\theta_j$ implies that the MCMC is more likely to move between $\theta_i$ and $\theta_j$, and vice versa. This leads us to the first ideal property for a proximity-map.

\begin{idProp}{}
    S
    If $\mathcal{M}$ is more likely to move from $\theta_i$ to $\theta_j$ than it is to move from $\theta_i$ to $\theta_k$, then 

    $$|\mathcal{P}(\theta_i) - \mathcal{P}(\theta_j)| \leq |\mathcal{P}(\theta_i) - \mathcal{P}(\theta_k)|.$$
\end{idProp}

Choosing a proximity-map that perfectly reflects Ideal Property 1 will ensure that the transformed MCMC draws (ie. $\mathcal{P}(\mathcal{C})$) behave in the way expected by standard univariate diagnostics. Because the parameter space $\Theta$ may have a large number of dimensions, it is frequently impossible to choose a $\mathcal{P}$ for which Ideal Property 1 is always true. Instead, this concept should serve as a guideline when selecting a proximity-map, with options that more closely align with Ideal Property 1 considered preferable.

The second ideal property involves the computational complexity of $\mathcal{P}$. Because we are transforming each draw from $\mathcal{M}$, the computational time required to use $\mathcal{P}$ is at least $\mathcal{O}(N)$ where $N \leq nK$ is the number of \textit{unique} draws from $\mathcal{M}$. We can, of course, choose a complex $\mathcal{P}$, possibly with the goal of adhering more closely to Ideal Property 1. However, if $N$ is large, this can easily result in prohibitively long computational times, simply to calculate diagnostics. We give Ideal Property 2 here, noting that `reasonable' is entirely subjective and may vary drastically depending on the investigator and the computing power available. 

\begin{idProp}{}
    S
    The computational complexity required to calculate $\mathcal{P}$ for the $N$ unique draws sampled using $\mathcal{M}$ allows for a `reasonable' computation time. 
\end{idProp}

Unfortunately, Ideal Property 1 and Ideal Property 2 frequently conflict. A proximity-map that more accurately reflects Ideal Property 1 may require more computational time, and a faster proximity-map may produce less accurate results leading to missed non-convergence diagnoses. Therefore, selecting a proximity-map is ultimately a trade-off based on investigator knowledge and preference. 

A few final considerations when choosing $\mathcal{P}$ involve additional choices that an investigator must make during the selection process. First, some proximity-maps require the definition of a function $f:\Theta \rightarrow \mathbb{R}$ or pairwise distance $d:\Theta \times \Theta \rightarrow \mathbb{R}$ which must also be chosen carefully. Additionally, the definitions of several of the proximity-maps we suggest below depend on either the samples or the order of samples drawn in $\mathcal{C}$, which an investigator should consider. Any of these choices can have an impact on the success of $\mathcal{P}$ as a convergence diagnostic, and must also be considered carefully. See Sections \ref{subsec: proxMaps} and \ref{sec: workingExamples} for more concrete examples of these decisions. 

Table \ref{tab: proxMaps} gives a brief overview of several classes of proximity-maps that may be used, although this list is by no means exhaustive. In Section \ref{subsec: proxMaps} we explore each of these classes in more depth, giving detailed definitions and in Section \ref{sec: workingExamples} we give examples of proximity-map usage while also discussing our reasoning for selecting each proximity-map. 

%A proximity-map that depends on $\mathcal{C}$ is one in which the definition of $\mathcal{P}$ can change with a different set of MCMC samples, and a proximity-map that requires $d$ is dependent on the definition of a pairwise distance. 

\begin{table}[t]
    \centering
    \begin{tabular}{|c|L|K|J|c|}
    \hline
        Class & Summary & \mline{Depends on $\mathcal{C}$} & \mline{Requires $d$} & Complexity\\
        \hline\hline
        Function & \mline{$\mathcal{P}(\theta) = f(\theta)$ for some function $f$}& & & $\mathcal{O}(N)$\\
        \hline
        Reference & \mline{$\mathcal{P}(\theta) = d(\theta, \theta_0)$ for a selected reference point $\theta_0$} & & \checkmark & $\mathcal{O}(N)$ \\
        \hline
        Nearest Neighbor & \mline{$\mathcal{P}(\theta)$ algorithmically defined using Nearest Neighbor Algorithm and distance $d$} & \checkmark & \checkmark & $\mathcal{O}\big(N^2\big)$\\ 
        \hline
        
    \end{tabular}
    \caption{Classes of proximity-maps with summary descriptions, requirements (depends on MCMC samples $\mathcal{C}$ or requires pairwise distance function $d$), and complexity}
    \label{tab: proxMaps}
\end{table}

\subsection{Classes of Proximity-Maps}\label{subsec: proxMaps}

In this section we give details on the classes of proximity-maps listed in Table \ref{tab: proxMaps}. For each we provide a detailed definition and reference examples where possible. 

\subsubsection{Function-Based Proximity-Maps}

We refer to our first class of proximity-maps as `function-based'. This class of proximity-maps simply sets $\mathcal{P}(\theta) = f(\theta)$ for all $\theta \in \Theta$ and for some pre-specified function $f$. In this case, $\mathcal{P}$ is generally computationally efficient (assuming $f$ is not overly computationally-burdensome) and can even be calculated while the MCMC algorithm is running. 

The most common examples of diagnostics derived from function-based proximity-maps are the likelihood-based diagnostics mentioned in Section \ref{subSec: liklihood}. In this case $f(\theta)$ is simply the likelihood (or score, log-likelihood, posterior, etc.) evaluated at $\theta$, and diagnostics are calculated using the transformed values. Examples of likelihood-based diagnostics may be found frequently in MCMC literature \cite{friedman2003being, grzegorczyk2008improving}. 

While function-based proximity-maps are generally not too computationally complex, the difficulty lies in choosing a function $f$ that may adequately satisfy Ideal Property 1. If $f$ does not properly account for the movement of the MCMC algorithm, the derived diagnostics may be misleading and fail to identify non-convergence. The other two classes of proximity-maps all seek to address this shortcoming by basing the proximity-map on a pairwise distance function $d$ that is chosen to reflect the movement of $\mathcal{M}$. We examine these next. 

\subsubsection{Reference-Based Proximity-Maps}

The reference-based family of proximity-maps depend on two choices, the choice of pairwise distance function $d$, and a reference point $\theta_0\in\Theta$. In this case, the proximity-map is defined by $\mathcal{P}(\theta) = d(\theta, \theta_0).$ 

Reference-based diagnostics have important advantages, including low computational-complexity and a flexible definition that may be tailored to specific models or algorithms, and are becoming more popular in Bayesian phylogenetic-tree analysis \cite{lanfear2016estimating, magee2024trustworthy} and entity-resolution \cite{aleshin2024convergence}. In these examples, investigators carefully select a pairwise distance function relevant to the problem formulation, and utilize the transformed draws in standard diagnostic calculations. (For further discussion on choosing $d$, see Section \ref{subSec: distances}.)

Unfortunately, the choice of a reference point $\theta_0$ can drastically change the resulting diagnostic information. Magee et al. discuss this issue in the context of phylogenetic trees, demonstrating that the choice $\theta_0$ can have a profound impact on the calculated ESS, and suggesting several approaches to solve the issue \cite{magee2024trustworthy}. Fundamentally, the issue with a reference-based proximity-map is that $|\mathcal{P}(\theta_i) - \mathcal{P}(\theta_j)| = |d(\theta_i, \theta_0) - d(\theta_j, \theta_0)|$ does not include the term $d(\theta_i, \theta_j)$ and can therefore stray very far from Ideal Property 1, even if the pairwise distance is carefully defined. 

The final class of proximity-maps we present attempts to improve upon this issue, albeit with a loss in computational efficiency. 

\subsubsection{Nearest Neighbor Proximity-Maps}

Nearest Neighbor (NN) proximity-maps are based on the Nearest Neighbor algorithm, a greedy graph-theoretic algorithm that provides a solution to the well-known Traveling Salesman problem \cite{reinelt2003traveling}. Defining this class of proximity-maps requires a thorough explanation of the NN problem formulation and algorithm, which we now present.

Consider a space $\Theta$ with an associated pairwise distance function $d$. Given the subset $\{\theta_1, \dots, \theta_N\} \subseteq \Theta$, the Nearest Neighbor algorithm produces an ordering of $\{\theta_1, \dots, \theta_N\}$ by picking a random value $\theta_i = \phi_0$ and then, step by step, moving to the next closest value (in terms of the distance function $d$) that has not been previously selected, until ending back at $\phi_0$. This is outlined in psuedo-code in Algorithm \ref{alg: NearestNeighbor}.

\begin{algorithm}[t]
\caption{The Nearest Neighbor algorithm}\label{alg: NearestNeighbor}
\begin{algorithmic}
\State \textbf{Input:} $\{\theta_1, \dots, \theta_N\},$ all unique, and a distance $d$. 
\State Set: $\{\theta_1, \dots, \theta_N\} = S$
\State Choose: $\phi_0 = \theta_i \in S$
\State Set: $S_0 = S \setminus \{\theta_i\}$

\For{$t \in 1,\dots, N-1$}
\State Set: $\phi_t = \argmin_{\theta \in S_t} d(\phi_{t-1}, \theta)$
\State Set: $S_t = S_{t-1} \setminus \{\phi_t\}$
\EndFor

\Return $(\phi_0, \phi_1, \dots, \phi_{N-1}, \phi_0)$
\end{algorithmic}
\end{algorithm}

Thus, the Nearest Neighbor algorithm can be used to take the set of unique sampled values from $\mathcal{M}$ and produce a cyclic ordering of those samples that respects the distance $d$. However, in order to define a proximity-map, we require an additional step.  

The output provided by Algorithm \ref{alg: NearestNeighbor} is a cycle, ending where it began, and not a true ordering. As we plan to map sampled values to the real-line respecting our distance-based ordering, we need to identify a place to `cut' our cycle (the so-called \textit{cut-point}) so that we have a beginning and ending in the mapped values. 

One approach to identifying the cut-point would be to find $\phi_m$ for which $d(\phi_{m-1}, \phi_m)$ is the largest, setting $\phi_{m}$ to be the beginning and $\phi_{m-1}$ to be the ending of our mapping. However, this assumes that the pairwise distance $d$ accurately reflects the movement of the MCMC sampler, which may not be the case. As such, we suggest a cut-point selection method that is adaptive and depends on the samples $\mathcal{C}$. 

Intuitively, we approach the cut-point selection as follows, with full definitions below. First, we will define a proximity-map, denoted $P^{(m)}$ with $m = 0, \dots, N-1$, for each possible cut-point in the cycle derived from Algorithm \ref{alg: NearestNeighbor}. Specifically, $P^{(m)}$ will treat the cut-point as occurring between $\phi_{m-1}$ and $\phi_m$. Next, consider drawing a traceplot of all chains $C_1, \dots, C_K$ after applying the proximity-map $P^{(m)}.$ We define a construct $D_k^{(m)}$ which simply measures the vertical distance traveled by chain $C_k$ under the mapping $P^{(m)}$. A (relatively) large value of $D_k^{(m)}$ implies that the chain travels from near the top to near the bottom of the traceplot's $Y$-axis frequently. If a single chain travels in this way it can obscure information about the other chains on the traceplot, as well as lower the accuracy of other univariate multi-chain diagnostics. Thus, we select the cut-point to correspond to the proximity-map $P^{(m)}$ that minimizes the sum of $D_{k}^{(m)}$ over all chains. This is provided in full detail now below. 

Let $\phi^{(m)} = (\phi_m, \phi_{m+1}, \dots, \phi_{N-1}, \phi_0, \dots, \phi_{m-1})$ be the ordering of $\{\theta_1, \dots, \theta_N\}$ created by cutting the cycle provided by the NN algorithm between $\phi_{m-1}$ and $\phi_m$, for $m\in \{0, \dots, N-1\}$. We will define a function $P^{(m)}$ based on $\phi^{(m)}$ for all $m$ and choose the function that minimizes a target distance defined below. Intuitively, we define $P^{(m)}(\phi_r)$ to be the distance traveled if moving from $\phi_m$ to $\phi_r$ point-by-point along the sequence $\phi^{(m)}$. Specifically, we recursively define $P^{(m)}:\{\theta_1, \dots, \theta_N\} \rightarrow \mathbb{R}$ by

\[
P^{(m)}(\phi_{(m+k+1)(\text{mod}N)}) = P^{(m)}(\phi_{(m+k)(\text{mod}N)}) + d(\phi_{(m+k)(\text{mod}N)}, \phi_{(m+k+1)(\text{mod}N)})
\]

\noindent for all $k \in \{0, 1, \dots, N-2\}$, and $P^{(m)}(\phi_m) = 0.$ As an example we have written out $P^{(2)}$ here:

\begin{align*}
    P^{(2)}(\phi_2) &= 0 \\
    P^{(2)}(\phi_{3}) &= d(\phi_2, \phi_{3}) \\
    P^{(2)}(\phi_{4}) &= d(\phi_2, \phi_{3}) + d(\phi_{3}, \phi_{4}) \\
    &\dots \\
    P^{(2)}(\phi_0) &= \sum_{i=3}^{N-1}d(\phi_{i-1}, \phi_{i}) + d(\phi_{N-1}, \phi_0) \\
    P^{(2)}(\phi_{1}) &= \sum_{i=3}^{N-1}d(\phi_{i-1}, \phi_{i}) + d(\phi_{N-1}, \phi_0) + d(\phi_{0}, \phi_{1}). \\
\end{align*}

Now, as defined above, let $C_k = (\theta_{1k}, \dots, \theta_{nk})$ be the $k$th chain of length $n$ drawn from $\mathcal{M}$. Define 

$$D_k^{(m)} = \sum_{i=2}^n \big|P^{(m)}(\theta_{ik}) - P^{(m)}(\theta_{(i-1)k})\big|.$$

%\noindent While the definition of $D_k^{(m)}$ is quite involved, it is actually a rather intuitive construct. Consider drawing a traceplot of all chains $C_1, \dots, C_K$ after applying the proximity-map $P^{(m)}.$ $D_k^{(m)}$ simply measures the vertical distance traveled by chain $C_k$ under this mapping. A (relatively) large value of $D_k^{(m)}$ implies that the chain travels from near the top to near the bottom of the traceplot's $Y$-axis frequently. If a single chain travels in this way it can obscure information about the other chains on the traceplot, as well as lower the accuracy of other univariate multi-chain diagnostics. Thus, we select the cut-point to be the value $\phi_m$ that minimizes the sum of $D_{k}^{(m)}$ over all chains and we define the Nearest Neighbor proximity-map as

\noindent Finally, we define the Nearest Neighbor proximity-map as

\[
\mathcal{P} = \argmin_{P^{(m)}|m \in \{0,\dots,N-1\}} \sum_{k=1}^KD_{k}^{(m)}.
\]

\noindent The psuedo-code for generating the Nearest Neighbor proximity-map is described in Algorithm \ref{alg: NNProx}.

\begin{algorithm}
\caption{Nearest Neighbor Proximity-map Generation}\label{alg: NNProx}
\begin{algorithmic}
\State \textbf{Input:} All chains $\mathcal{C}$ drawn from $\mathcal{M}$, and a distance $d$.
\State Identify: $\{\theta_1, \dots, \theta_N\}$, the set of unique values in $\mathcal{C}$.
\State Evaluate: Algorithm \ref{alg: NearestNeighbor} with $(\theta_1, \dots, \theta_N)$ and $d$. Return solution $(\phi_0, \dots, \phi_{N-1}, \phi_0)$.
\For{$m \in 0, \dots, N-1$}
\State Set: $D^{(m)} = \sum_{i=1}^k D^{(m)}_{i}$ 
\EndFor

\State Choose: $P^{NN}_d = f^{(s)}$ such that $D^{(s)} = \min_m D^{(m)}$

\Return $P^{NN}_d:(\theta_1, \dots, \theta_N) \rightarrow \mathbb{R}$.
\end{algorithmic}
\end{algorithm}

A Nearest Neighbor proximity-map has a key advantage over the other classes of proximity-maps discussed here. Because it is based on pairwise distances between all unique draws and not just a single reference point, it frequently captures more of the true sample space in the final mapping, and can adhere more closely to Ideal Property~1. 

However, the Nearest Neighbor proximity-map is \textit{significantly} more computationally-intensive than either of the other classes, particularly when the number of unique drawn values $N$ is very large. Additionally, as is well-documented, the Nearest Neighbor algorithm is sensitive to the choice of the starting point \cite{reinelt2003traveling}, an important issue of which to be aware. % This doesn't negatively affect the Nearest Neighbor proximity-map as badly as the choice of $\theta_0$ for a reference-based proximity-map, but it is an important issue of which to be aware.

\subsection{Considerations for Choosing a Pairwise Distance Function}\label{subSec: distances}

The process of selecting a useful pairwise distance is highly dependent on the MCMC algorithm or model, and the particular aspect of convergence that an investigator is interested in exploring. Therefore, we do not provide a comprehensive list of possible distance functions to choose from, as such a list would be beyond the scope of this paper. Instead we discuss guidelines for selecting $d$ and two examples of this process in Section \ref{sec: workingExamples}.

The first consideration when choosing a pairwise distance $d$ is to ensure that it aligns with the Ideal Properties for proximity-maps. For an ideal function $d$ we would have that $d(\theta_i, \theta_j) \leq d(\theta_i, \theta_k)$ whenever $\mathcal{M}$ is more likely to move from $\theta_i$ to $\theta_j$ and less likely to move from $\theta_i$ to $\theta_k$. Additionally, as some proximity-maps require $d$ to be calculated up to $N^2$ times, ideally $d$ is computationally simple and fast. 

An investigator may also consider limiting the scope of the distance function to focus on the convergence of a particular portion of the model, such as only including parameters involved in covariate selection in the definition of $d$. In this case, the resulting convergence diagnostics will summarize convergence on a portion of the model, similar to examining univariate convergence diagnostics in a multivariate model. This practice, as we will demonstrate below, can lead to illuminating answers about model performance. 

Finally, it is generally helpful, if not strictly necessary, for the pairwise distance function to be easily interpretable. 

%Convergence diagnostics are much less convincing when they are presented with a long-winded and confusing explanation about their derivation. 

We also suggest that, when presented with several possibly useful pairwise distance functions, an investigator test all options available as they may provide different and complementary information about the behavior of the MCMC algorithm. 

\section{Working Examples}\label{sec: workingExamples}

Here we present examples of two MCMC-based situations in which we use our proximity-map framework to easily develop useful convergence diagnostics and gain insight into the behavior of an MCMC algorithm. The first example presents a case involving an MCMC algorithm for a Dirichlet process mixture model (DPMM) \cite{neal2000markov}, and the second discusses an MCMC algorithm for a Bayesian Kernel Machine Regression (BKMR) model \cite{bobb2015bayesian}. 

\subsection{Dirichlet Process Mixture Model}

The Dirichlet process mixture model (DPMM) may be used to cluster observations into latent groups, and is unique in that its construction does not require the number of groups to be specified \textit{a priori}. The algorithm and its theoretical justification were first introduced approximately 50 years ago in a series of papers by Antoniak \cite{antoniak1974} and Ferguson \cite{ferguson1973, ferguson1983}. While the use of this Bayesian nonparametric approach is highly flexible and provides uncertainty quantification regarding the number of latent groups, the algorithms used to fit the DPMM are computationally intensive \cite{neal2000markov}. 

%The Dirichlet process prior which underlies the DPMM may be described using the Chinese restaurant process \cite{gonen}.

The primary difficulty encountered by standard convergence diagnostics in a DPMM is the varying dimensionality over the course of the MCMC iterations. As discussed in Section \ref{sec: limitations}, the potential for frequent change in the dimension of the parameter space makes it impossible to examine convergence diagnostics for particular parameters or sets of parameters. Thus, in order to make standard traceplots or calculate the values of standard diagnostic measures, one must condition on the number of latent components found at each MCMC iteration, and evaluate diagnostics on those results separately. These MCMC iterations are not necessarily contiguous after conditioning, so this results in an incomplete or potentially misleading picture of the path of the MCMC.

To demonstrate the utility of the proximity-map based diagnostics introduced in this paper, we simulated data from a mixture of three Gaussian densities, denoted $\mathcal{X}_D$, and used it to fit a DPMM. The true component means are $(-20, 20)$, $(0,0)$, and $(20, -20)$, variances are all $10\cdot I_2$, and the mixing proportions are $(0.4, 0.3, 0.3)$. We fit a DPMM with a diagonal covariance structure, setting a normal prior on the mean parameters $\mu_k \sim N(\mu_0, \sigma^2_0)$ and an inverse gamma prior on the diagonal variances $\sigma^2_k \sim IG(a,b)$. Using Algorithm 2 from Neal \cite{neal2000markov}, we learned parameters $\mu_1, \dots, \mu_K$, and $\sigma^2_1, \dots, \sigma^2_K$ for latent groups $k = 1, \dots, K$ (noting again that $K$ changes between iterations). We ran three chains with this MCMC sampler, for 5,000 iterations each. 

\begin{figure}[t]
    \centering
    \begin{subfigure}{.5\textwidth}
    \centering
    \includegraphics[width = \textwidth]{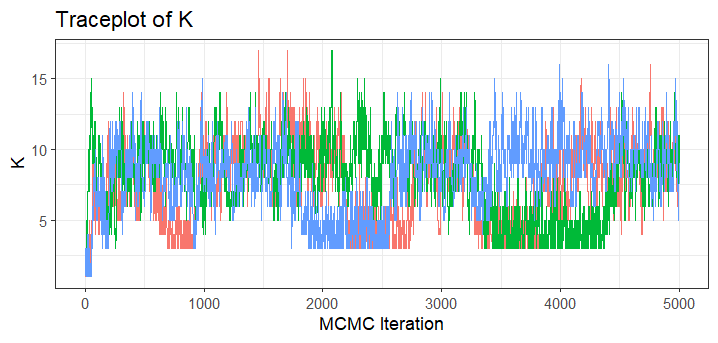}
    \caption{Traceplot for $K$.}
    \label{subfig: DPMMK}
    \end{subfigure}%
    \begin{subfigure}{.5\textwidth}
    \centering
    \includegraphics[width = \textwidth]{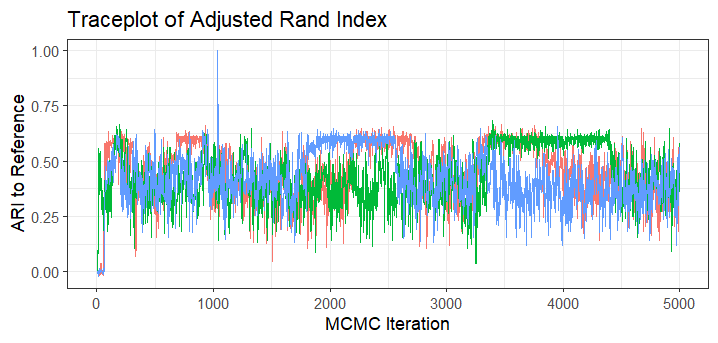}
    \caption{Traceplot for ARI.}
    \label{subfig: DPMMari}
\end{subfigure}
\caption{Traceplots for $K$ and ARI for the DPMM model fit on the simulated data using Algorithm 2 from Neal \cite{neal2000markov}.}
\label{fig: DPMM}
\end{figure}

A typical approach to diagnosing convergence in a DPMM might begin with examining the traceplot for $K$, the number of non-empty latent groups at each iteration. Figure \ref{subfig: DPMMK} presents this traceplot. This plot shows that the algorithm is mixing relatively well across chains on the parameter $K$, jumping between two modes around $K = 4$ and $K = 10$, and still should be run for much longer to achieve full mixing. However, the parameter $K$ only represents a small fraction of the information given by the parameters learned at each iteration of the sampler, and additional investigation is warranted. Notably, the group assignments for each observation are much more informative than $K$, but more difficult to summarize and compare between consecutive MCMC iterations. Thus, we selected two proximity-maps to use for additional diagnostics, one focused on the clustering portion of the model, and the other focused on the density estimation. 

For the first proximity-map, we targeted the clustering portion of the MCMC algorithm. At each MCMC iteration $t$, we extract a categorical vector $X_t$ assigning each observation to a different cluster. Since the number of observations in a DPMM is constant, the vector $X_t$ has the same dimension for all $t$, but contains a varying number of unique values, depending on the number of groups at iteration $t$, $K_t$. Additionally, the vector $X_t$ can suffer from a `label-switching' problem \cite{stephens2000dealing}. That is, for $t \neq s$ even if the clusterings represented by $X_t$ and $X_s$ are identical, the values assigned to each observation in $X_t$ may be different from the values assigned to $X_s$. In order to avoid this issue, we randomly selected a clustering from among those sampled, $X_{t_0}$ and utilized a reference-based proximity-map with the Adjusted Rand Index (ARI) as the pairwise distance \cite{rand, adjrand}. This works particularly well, as the ARI is a label-agnostic measure of classification performance that summarizes the pairwise concordance between two different clusterings and provides insight into the amount that the sampler changes the clustering across iterations. 

Figure \ref{subfig: DPMMari} gives the traceplot after utilizing the reference-based proximity-map and the ARI. This traceplot provides evidence that each chain successfully moves around the clustering space, and that the three chains are mixing well together. As with Figure \ref{subfig: DPMMK}, there is evidence that each chain is jumping between two distinct modes, and it is encouraging to see that the two traceplots in Figure \ref{fig: DPMM} agree on when each chain is in which mode. Based on these traceplots alone, however, it does seem that the MCMC algorithm could be run for significantly longer to ensure that the model converges successfully. 

For our second convergence measure for the same set of draws, we targeted the density estimation portion of the MCMC algorithm used to fit the DPMM. Within a given MCMC iteration, each observation is assigned to a cluster represented by a Gaussian distribution. Thus, in each iteration we may extract a vector of z-scores, $Z_t$, that gives the standardized distance from each observation to its assigned cluster, using cluster specific means and variances. We developed convergence diagnostics using the Nearest Neighbor proximity-map with the pairwise distance $d(Z_t, Z_s)$ defined as the Euclidean distance between the vectors $Z_t$ and $Z_s$. 

The goal of the Nearest Neighbor proximity-map with the z-scores was to quantify the mixing using ESS and PSRF measures, in case the algorithm was failing to converge in the density estimation space. Using this approach we identified the ESS as approximately 400 per chain, and the PSRF as 1.008. These results suggest that the MCMC is successfully mixing across chains, but needs to be run for longer to ensure good convergence within each chain and therefore a sufficient sample size. 

While these additional diagnostics gave similar information to diagnostics based solely on the parameter $K$ in this scenario, this may not always be the case. In parameter-rich models like the DPMM, we would like to incorporate information from all parameters into our diagnosis of whether sufficient mixing, and eventually, convergence, have been achieved.

\subsection{Bayesian Kernel Machine Regression with SELF Data}

We now detail a process of improving the probit-BKMR MCMC algorithm after multiple investigators, utilizing the the algorithm as available from the \texttt{bkmr} package, reported that the algorithm failed to converge. Our proximity-map diagnostic approach was instrumental in quickly and efficiently identifying the issue. The improvements are available in the probit-BKMR MCMC algorithm on the \texttt{bkmr} GitHub repository as of November 14, 2024. In the present example, we utilize data from the Study of Environment, Lifestyle, and Fibroids (SELF) cohort.

SELF is a prospective cohort designed to identify risk factors for the development of uterine fibroids \cite{baird2015prospective}. Between 2010-2012, a total of 1,693 reproductive-aged women (age 23-35 years) from the Detroit, MI area who self-identified as Black or African American were enrolled. Eligible participants had an intact uterus, no previous diagnosis of fibroids, and no history of cancer or autoimmune disorders that required medication. Participants were screened for fibroids with ultrasound at enrollment and at follow-up visits that occurred approximately every 20 months for 10 years. At the baseline visit, participants completed questionnaires to provide information on sociodemographics, reproductive history, and other behavioral and exposure data. Participants provided non-fasting venous whole blood and spot urine specimens at the baseline visit. Blood and urine samples at baseline were analyzed for metals, persistent endocrine disrupting chemicals (EDCs, i.e., PCBs, PBDEs, PBBs, PFAS, organochlorine pesticides), and non-persistent EDCs (i.e., phthalates, phenols, parabens, triclocarban). SELF investigators have published several studies examining the effects of multi-pollutant mixtures on fibroid development (see for example: \cite{wesselink2021urinary, fruh2024non, geller2025prospective}). 

The complex interactions inherent in estimating the effects of multi-pollutant exposures require a unique analytical approach. The Bayesian kernel machine regression (BKMR) model provides an excellent approach to these complexities by modeling the target outcome by:

\[
Y_i = h(z_{i1}, \dots, z_{iM})  + X_i\beta + \epsilon_i,
\]

\noindent where $Y_i$ is a continuous outcome for individual $i = 1, \dots, n$, $z_{i1}, \dots, z_{iM}$ are the exposure values of interest, $h$ is an exposure-response function represented using kernel machine regression, $X_i$ is a vector of baseline covariates with parameters $\beta$, and $\epsilon_i$ is a normal error term \cite{bobb2015bayesian}. The use of the kernel machine representation of $h$ allows the multi-pollutant exposure effects to be estimated non-linearly and non-additively, and is operationally accomplished in a Bayesian framework by setting $Y_i = h_i + X_i\beta + \epsilon_i$ and assuming

\[
h = (h_1, \dots, h_n) \sim N(0, \lambda\mathbf{K}),
\]

\noindent where $\mathbf{K}$ is called the \textit{kernel matrix}, and is based on the exposure values $z_{ij}$ (see \cite{bobb2015bayesian} for details). 

Additionally, the inclusion of component-wise variable selection of the exposures in the BKMR model allows for the identification of specific pollutants from the mixture that are driving the observed health effects. The variable selection is represented using parameters $\delta_1, \dots, \delta_M$ where $\delta_j = 1$ if exposure $z_{ij}$ is included in the model, and $\delta_j = 0$ otherwise. The model is fit using an MCMC with a combination of Gibbs and Metropolis-Hastings steps.

In the case that the target outcome is binary, the BKMR model may be adjusted to a probit model using the latent normal formulation:

\[
Y_i^* = h(z_{i1}, \dots, z_{iM})  + X_i\beta + e_i
\]

\noindent where $e_i$ is a standard normal error (the variance is set to 1) and $Y_i^*$ expresses a latent quantity controlling $Y_i$ by $Y_i = 1$ if $Y_i^* > 0$ and $Y_i = 0$ if $Y_i^* \leq 0$ \cite{albert1993bayesian}. The kernel machine representation, variable selection and MCMC fitting remain almost entirely unchanged (see \cite{bobb2018statistical} for details).

For the initial analysis we fit a probit-BKMR model with component-wise variable selection using the \texttt{bkmr} and \texttt{bkmrhat} R packages available on CRAN \cite{bobb2022bkmr, kiel2022bkmrhat}. The model was run on 3 separate chains with 5,000 MCMC iterations each. The target outcome was the incidence of uterine fibroids detected at the first follow-up visit, and the multi-pollutant exposure mixture was a set of 17 whole-blood metal concentrations. Analysis details and results based on well-mixed (both within- and across-chains) MCMC runs are presented in full in \cite{geller2025prospective}. In the following we discuss a particular random seed and subset of the SELF data (800 randomly selected individuals) for which the MCMC failed to converge successfully and which was not presented in \cite{geller2025prospective}.  %Details on inclusion criteria for this analysis, tables for analysis cohort characteristics, and a complete list of exposures and baseline covariates included in the model may be found in the Supplementary Material.

\begin{figure}[t]
    \centering
    \begin{subfigure}{.5\textwidth}
    \centering
    \includegraphics[width = \textwidth]{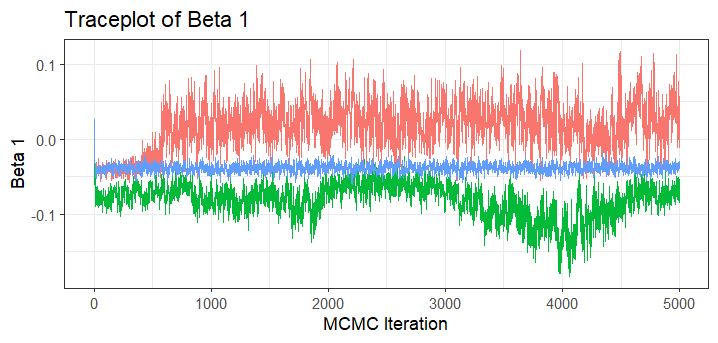}
    \caption{Traceplot for $\beta_1$.}
    \label{subfig: SELFBeta1}
    \end{subfigure}%
    \begin{subfigure}{.5\textwidth}
    \centering
    \includegraphics[width = \textwidth]{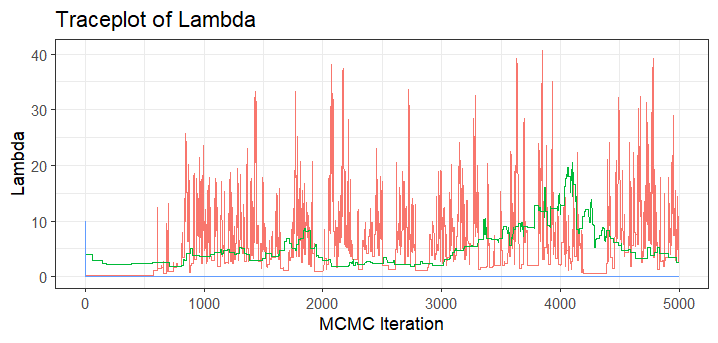}
    \caption{Traceplot for $\lambda$.}
    \label{subfig: SELFLambda}
\end{subfigure}
\caption{Traceplots for $\beta_1$ and $\lambda$ for the original BKMR model fit on a subset of the SELF data using \texttt{bkmr} as available on CRAN.}
\label{fig: SELFOrig}
\end{figure}

As a first step in examining the convergence of this MCMC run, we look at the traceplot of $\beta_1$, the parameter for the baseline covariate, participant age. As demonstrated in Figure \ref{subfig: SELFBeta1}, each chain is mixing fairly well \textit{within chain}, but none of the three are mixing well \textit{across chains}. Obviously, these results show that the BKMR model has not converged and should not be used for inference. However, in the interest of diagnosing and preventing further issues in the MCMC fit, further exploration is warranted.

Figure \ref{subfig: SELFLambda} gives a traceplot of $\lambda$, the kernel machine regression variance parameter for $h$, and begins to reveal exactly \textit{why} the MCMC is failing to converge. We can clearly see three distinct patterns of behavior exhibited by the three chains in Figure \ref{subfig: SELFLambda}. The chain represented in green appears to slowly move around the parameter space, while the chain in red jumps aggressively and sporadically and the chain in blue does not actually move at all. 

\begin{figure}[t]
    \centering
    \begin{subfigure}{.5\textwidth}
    \centering
    \includegraphics[width = \textwidth]{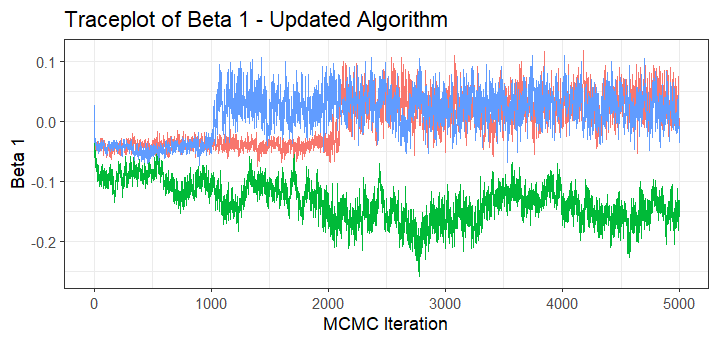}
    \caption{Traceplot for $\beta_1$.}
    \label{subfig: SELFBeta1Up}
    \end{subfigure}%
    \begin{subfigure}{.5\textwidth}
    \centering
    \includegraphics[width = \textwidth]{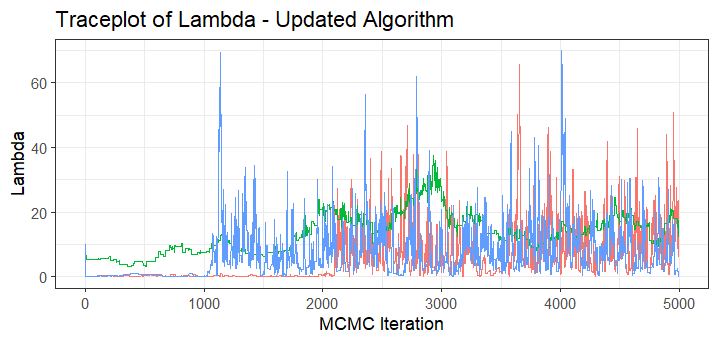}
    \caption{Traceplot for $\lambda$.}
    \label{subfig: SELFLambdaUp}
\end{subfigure}
\caption{Traceplots for $\beta_1$ and $\lambda$ for the updated BKMR model fit on the SELF data.}
\label{fig: SELFUpdated}
\end{figure}

Of the three possible behaviors, the behavior of the blue chain is the most concerning as it appears that the Metropolis-Hastings step here is unable to propose a single value that can be accepted. In order to mitigate this issue we modified the proposal distribution used in the Metropolis-Hastings step for $\lambda$, seeking to discourage the behavior shown by the blue chain. The resulting traceplots from this updated algorithm run on the same data are shown in Figure \ref{fig: SELFUpdated}.

As can be easily seen from Figure \ref{subfig: SELFBeta1Up}, the MCMC is still not converging across chains successfully. Additionally for $\lambda$, as seen in Figure \ref{subfig: SELFLambdaUp}, while all three chains are now actively moving around the parameter space, the blue and red chains are still exhibiting very different behavior from the green chain, with both sporadically jumping in the parameter space. We now investigate this behavior using a proximity-map based traceplot that uses information from multiple parameters simultaneously.

\begin{figure}[t]
    \centering
    \includegraphics[width=0.75\textwidth]{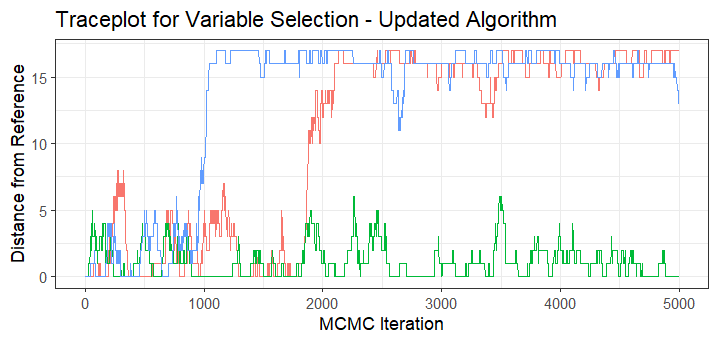}
    \caption{Proximity-map based traceplot of the variable selection parameters in the updated BKMR model fit on the SELF data.}
    \label{fig: SELFVarSel}
\end{figure}

Figure \ref{fig: SELFVarSel} gives a proximity-map based traceplot of the variable selection components, $\Delta = (\delta_1, \dots, \delta_{17})$, involved in the run of the updated probit-BKMR algorithm on the SELF data. The proximity-map is reference-based with reference $\Delta_0 = (\delta_1,\dots, \delta_{17}) = (1, \dots, 1),$ and utilizes the Euclidean distance for $d(\Delta_0, \Delta_t)$. The traceplot was created using our R package \texttt{genMCMCDiag} \cite{duttweiler2024genMCMC}. We selected this proximity-map in order to examine the behavior of the MCMC variable selection process \textit{as a whole}, rather than just looking at individual parameters. The information we glean from Figure \ref{fig: SELFVarSel} is very illuminating, and not available using standard convergence diagnostics, which only are able to examine one parameter at a time.

Figure \ref{fig: SELFVarSel} demonstrates that the variable selection portion of the probit-BKMR model is influencing the behavior of the parameter $\lambda$. By comparing the traceplots in Figures \ref{subfig: SELFLambdaUp} and \ref{fig: SELFVarSel} we can see that the odd behavior exhibited for $\lambda$ in the red and blue chains occurs precisely when the algorithm removes most or all of the exposures from the model.

\begin{figure}[t]
    \centering
    \begin{subfigure}{.5\textwidth}
    \centering
    \includegraphics[width = \textwidth]{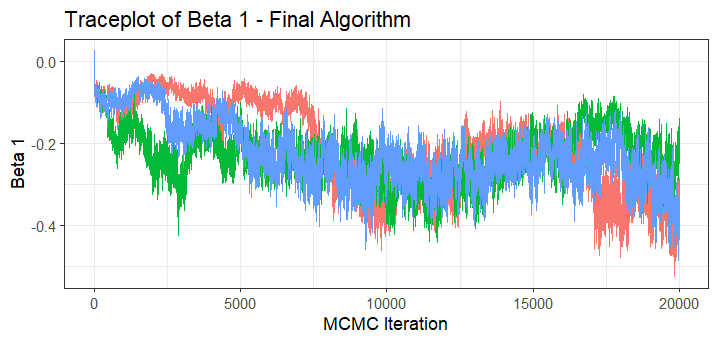}
    \caption{Traceplot for $\beta_1$.}
    \label{subfig: SELFBeta1Fin}
    \end{subfigure}%
    \begin{subfigure}{.5\textwidth}
    \centering
    \includegraphics[width = \textwidth]{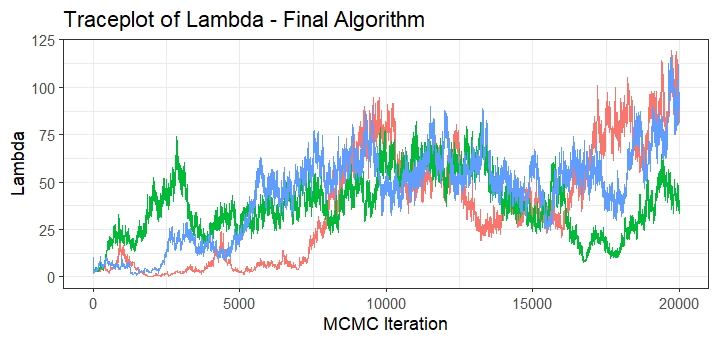}
    \caption{Traceplot for $\lambda$.}
    \label{subfig: SELFLambdaFin}
\end{subfigure}
\caption{Traceplots for $\beta_1$ and $\lambda$ for the final BKMR model fit on the SELF data.}
\label{fig: SELFfinal}
\end{figure}

Based on this additional information we made another change to the proposal step for $\lambda$ in the probit-BKMR algorithm, to encourage smaller, more regular movements even when $\lambda$ is very small and most or all of the exposures are included in the model. The traceplots for an MCMC run with 20,000 iterations based on this final algorithm adjustment are shown in Figure \ref{fig: SELFfinal}. As can be seen, the model still needs to run for a longer period of time to fully converge, but it is not getting stuck anywhere, and all three chains show improved mixing. This new adjustment that allows the probit-BKMR algorithm to consistently converge is incorporated into the development version of \texttt{bkmr} on GitHub as of November 14, 2024. 

\section{Discussion}\label{sec: discussion}

As demonstrated through the working examples, proximity-map based convergence diagnostics and traceplots can quickly and easily provide insight regarding MCMC convergence to investigators, even for complex models with many parameters and varying-dimensionality. The strengths of our suggested approach include flexibility and easy implementation on virtually any MCMC. Additionally, we have made an R package available on CRAN, \texttt{genMCMCDiag}, which implements the methods shown in this paper, providing simple access to convergence diagnostics \cite{duttweiler2024genMCMC}. However, it is important to note that the proximity-map approach suggested in this paper is very general and does not tailor to any particular class of MCMC methods. Because of this, we expect that convergence diagnostics customized for a specific algorithm will out-perform our approach in most situations, and may in some situations be accompanied with theoretical guarantees. Nonetheless, an important contribution of this work is that it gives researchers the ability to check for convergence when no standard or tailor-made convergence diagnostics are available. 

Future research directions include the development of improved classes of proximity-maps, research into further computational enhancements to the available methods, and publishing additional tutorials and guidance on identifying convergence issues with proximity-maps. Here our contribution is to enhance the usefulness of complex MCMC methods by streamlining the implementation of convergence diagnostics. We hope that access to these additional tools will encourage broader, thoughtful application of these powerful algorithms.

\noindent\textbf{Declaration of Competing Interest:}

\noindent There is no competing interest.

\vspace{.4cm}

\noindent\textbf{Acknowledgements:}

\vspace{.3cm}

\noindent The authors would like to thank SELF researchers and staff who designed and conducted the study, and participants who shared their biospecimens and their personal data. Dr. Serge Aleshin-Guendel provided helpful suggestions of additional citations and applications for our work, and Luke Rosamilia suggested the title. 

\vspace{.3cm}

\noindent Research reported in this publication was supported by the National Institute of Environmental Health Sciences of the National Institutes of Health (NIH), an agency of the U.S. Department of Health and Human Services, under award numbers T32ES007271, T32ES007142, P30ES000002, and P42030990, the Intramural Research Program of the NIH, National Institute of Environmental Health Sciences under award numbers ZIAES09013 and R01ES028235, and funds from the American Recovery and Reinvestment Act funds designated for NIH research. The content is solely the responsibility of the authors and does not necessarily represent the official views of the NIH.

\clearpage

\bibliographystyle{ba}
\bibliography{bibliography}

\end{document}